\documentclass[reprint,aps,prl,nobibnotes,superscriptaddress,english]{revtex4-1}
\usepackage[T1]{fontenc}
\usepackage[latin9]{inputenc}
\usepackage{float}
\usepackage{bm}
\usepackage{amsmath}
\usepackage{amssymb}
\usepackage{graphicx, xcolor}
\usepackage{babel}
\usepackage{hyperref,xr}

\begin{document}

\title{Optomagnonics in dispersive media: magnon-photon coupling enhancement at the epsilon-near-zero frequency}

\author{V. A. S. V. Bittencourt}
\affiliation{Max Planck Institute for the Science of Light, 91058 Erlangen, Germany}

\author{I. Liberal}
\affiliation{Electrical and Electronic Engineering Department, Institute of Smart Cities (ISC), Universidad P\'{u}blica de Navarra (UPNA), 31006 Pamplona, Spain}

\author{S. {Viola Kusminskiy}}
\affiliation{Max Planck Institute for the Science of Light, 91058 Erlangen, Germany}
\affiliation{Department of Physics, University Erlangen-N{\"u}rnberg, 91058 Erlangen, Germany} 


\begin{abstract}
Reaching strong light-matter coupling in solid-state systems has been long pursued for the implementation of scalable quantum devices. Here, we put forward a system based on a magnetized epsilon-near-zero (ENZ) medium, and we show that strong coupling between magnetic excitations (magnons) and light can be achieved close to the ENZ frequency due to a drastic enhancement of the magneto-optical response. We adopt a phenomenological approach to quantize the electromagnetic field inside a dispersive magnetic medium in order to obtain the frequency-dependent coupling between magnons and photons. We predict that, in the epsilon-near-zero regime, the single-magnon single-photon coupling can be comparable magnon frequency for small magnetic volumes. For state-of-the-art illustrative values, this would correspond to achieving the single-magnon strong coupling regime, where the coupling rate is larger than all the decay rates. Finally, we show that the non-linear energy spectrum intrinsic to this coupling regime can be probed via the characteristic multiple magnon sidebands in the photon power spectrum. 
\end{abstract}

\maketitle

In the strong light-matter coupling regime, matter excitations and photons are reversibly exchanged over timescales shorter than the typical decay rates \cite{Diaz_2019_Ultrastrong,Frisk_2019_ultrastrong}, allowing the implementation of quantum information protocols, such as quantum state transfer \cite{Kurizki_2015_quantum_technologies}. Reaching strong coupling in solid-state systems is particularly relevant for the design of scalable quantum devices, with semiconductors a notable example of strong coupling between light and solid-state excitations \cite{Bayer_2017_Terahertz,Cheng_2018_Charged_Polaron,Peraca_2020_ultrastrong}. Among the different platforms based on solid-state systems, magnetic dielectrics have recently emerged as promising materials to incorporate into hybrid quantum systems, in particular for magnon-mediated transduction \cite{Lachance_Quirion_2019_Hybrid_Quantum,Rameshti_2021_Cavity_Magnonics}. Whereas strong magnon-photon coupling in the microwave regime is by now routinely achieved \cite{Rameshti_2021_Cavity_Magnonics,Lachance_Quirion_2019_Hybrid_Quantum,Wang_2020_Dissipative_couplings,Awschalom_2021_Quantum_engineering}, state-of-the-art {\it opto}magnonic systems have reached couplings of $\sim50\:{\rm Hz}$ \citep{Haigh_2020_subpicoliter,Zhu_2020_Waveguide_cavity,Zhu_2021_inverse}, which are considerably smaller than the typical magnon and photon decay rates, $\sim\,{\rm MHz}$ and $\sim{\rm GHz}$ respectively. This severely limits applications in the quantum regime \citep{Hisatomi_2016_Biderectional_conversion,Zhu_2020_Waveguide_cavity}.

In this work, we theoretically show that the long-sought, sinngle-particle strong coupling regime between magnons and light can be reached in magnetized epsilon-near-zero (ENZ) media, where the permittivity $\varepsilon$ vanishes for a range of frequencies whereas absorption remains low. ENZ media belong to the broader class of near-zero-index (NZI) media, in which one or more of the constitutive parameters approach zero \citep{liberal_near_zero_2017,Engheta_2013_pursuing_near,kinsey_near_zero_index_2019}, and can be realized in homogeneous \citep{Naik_2013_Alternative,Caldwell_2013_low_loss} or structured media \citep{Campione_2015_Epsilon_near,Li_2015_On_chip, vulis_manipulating_the_2018}. In such systems non-linear effects and secondary responses, such as magneto-optical (MO) effects, can be greatly enhanced \citep{Caspani_2016_enhanced_nonlinear,reshef_nonlinear_optical_2019,kinsey_near_zero_index_2019}. ENZ systems have been proposed for designing compact optical isolators \citep{davoyan_optical_isolation_2013}, for engineering backscattering-protected propagation regimes for surface waves \citep{davoyan_theory_of_2013,Abbasi_oneway_2015}, and for the enhancement of the transverse MO Kerr effect \citep{Almpanis_controllingTMOKE_2020,giron_giant_enhancement_2017}.

\begin{figure}
\begin{centering}
\includegraphics[width=1\columnwidth]{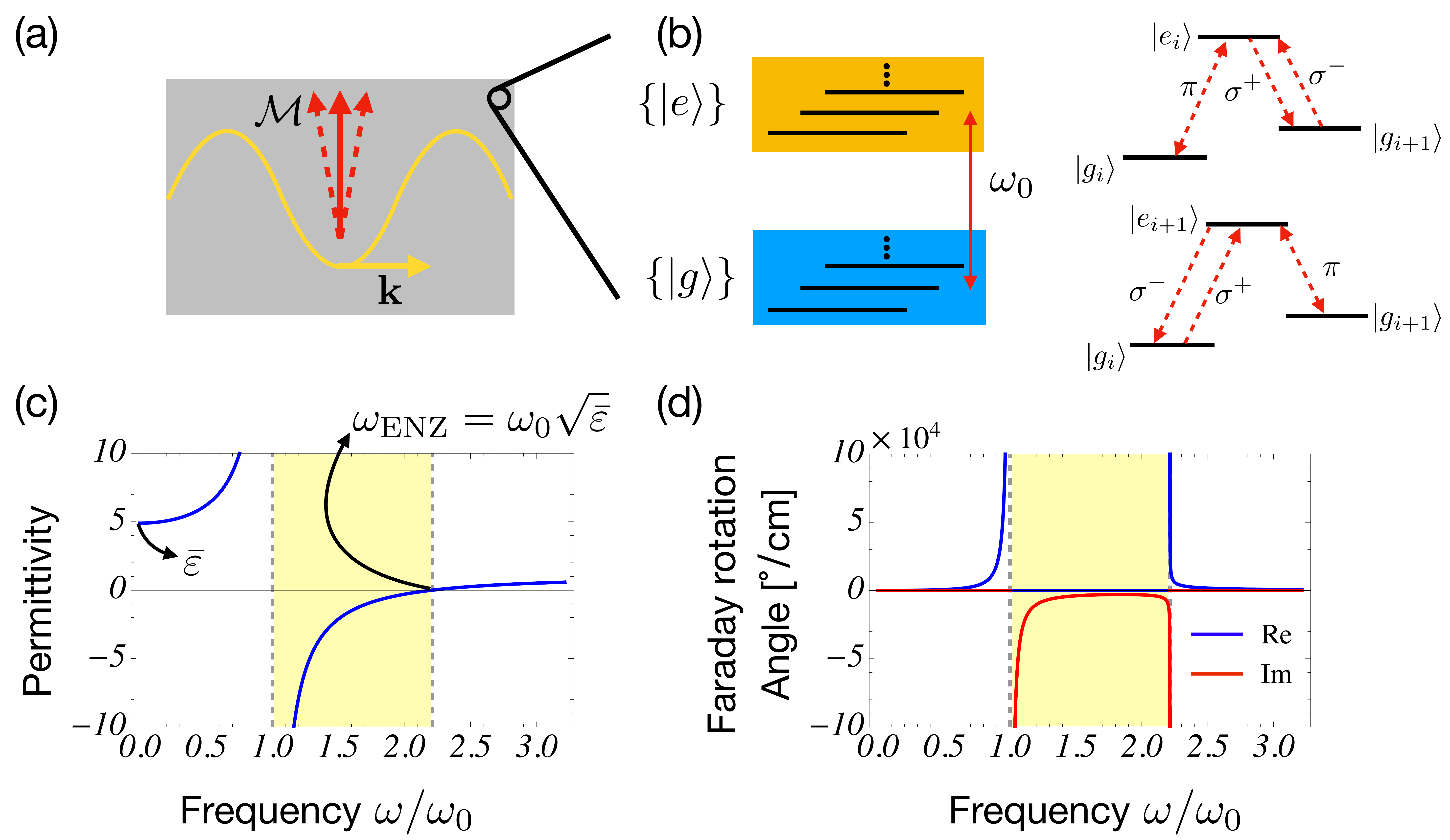}
\par\end{centering}
\caption{(a) Setup: a monochromatic electromagnetic wave propagates in a magnetized dielectric. The wave-vector is perpendicular to the magnetization (Voigt configuration), which exhibits small fluctuations around its saturation value; (b) Minimal microscopic model leading to the dispersion of the permittivity tensor: light drives electric-dipole transitions responsible for the Faraday effect and the material's optical response; (c) Permittivity (\ref{eq:Permittivity}) and (d) Faraday rotation angle (\ref{eq:FradayRotAngle-1}) as functions of the frequency for this model of dispersion. At the frequency $\omega_{{\rm ENZ}}$, $\varepsilon[\omega_{{\rm ENZ}}]=0$ and the Faraday rotation angle is enhanced. Illustrative parameters corresponding to a resonance of Yittrium Iron Garnet with ionic transition frequency $\omega_0= 2 \pi \times 600 \, \rm{THz}$.}
\label{Fig01}
\end{figure}

In what follows, we develop a Hamiltonian description for magneto-optical effects at the single-quanta level in dispersive magnetic dielectrics. We obtain the frequency-dependent magnon-photon coupling Hamiltonian
\begin{equation}
\hat{H}_{{\rm OM}}=i\hbar g[\omega_{c}]\hat{a}_{+}^{\dagger}\hat{a}_{-}(\hat{m}^{\dagger}+\hat{m})+{\rm h.c.},\label{eq:OMDispersionHamiltonian}
\end{equation}
between two degenerate optical modes with frequency $\omega_{c}$, where $\hat{a}_{\pm}^{(\dagger)}$ and $\hat{m}^{(\dagger)}$ are the bosonic annihilation (creation) photon and magnon operators respectively, and show that at the epsilon-near-zero frequency $\omega_{{\rm ENZ}}$
\begin{equation}
g[\omega_{{\rm ENZ}}]=\frac{\mathcal{M}_{{\rm ZPF}}}{\mathcal{M}_{S}}\omega_{{\rm ENZ}},\label{eq:GEZ}
\end{equation}
where $\mathcal{M}_{{\rm ZPF}}$ and $\mathcal{M}_{S}$ are, respectively, the zero-point magnetization fluctuations and the saturation magnetization. In a cavity setup, this coupling can be comparable to the magnon frequency, allowing to reach the single-magnon strong coupling regime, as we illustrate with a minimal model for the dispersion. This regime can be probed  by the light's power spectrum, which exhibits resonances characteristic of a non-linear energy spectrum generated by the strong single-magnon coupling.

We consider the framework of Fig.~\ref{Fig01} where light propagates in a dielectric ferromagnet in the Voigt configuration, perpendicular to the uniform magnetization $\bm{\mathcal{M}}$ of the medium. Our starting point for the quantization of the magneto-optical coupling is the energy density for a monochromatic field with a frequency $\omega_{c}$, $\bm{E}(\bm{r},t)={\rm Re}\left\{ \bm{E}^{(+)}(\bm{r},t)\right\} ={\rm Re}\left\{ \bm{E}^{(+)}(\bm{r})e^{-i\omega_{c}t}\right\} $, inside of a magnetic dielectric, which we assume to be homogeneous and such that absorption can be neglected in the frequency range of interest. Taking into account the dispersion of the medium's permittivity, the energy density reads \citep{Landau_electrodynamics_2009,Bittencourt_2021_ENZ}
\begin{equation}
\begin{aligned}
u_{\omega_{c}} & =\frac{\varepsilon_{0}}{4}\bm{E}^{(-)}(\bm{r},t)\cdot\partial_{\omega_{c}}\left(\omega_{c}\overleftrightarrow{\varepsilon}[\omega_{c},\bm{\mathcal{M}}]\right)\cdot\bm{E}^{(+)}(\bm{r},t)\\
 & \,\,+\frac{1}{4 \mu_{0}}\bm{B}^{(-)}(\bm{r},t)\cdot\bm{B}^{(+)}(\bm{r},t),
\end{aligned}
\end{equation}
where $\bm{B}(\bm{r},t)$ is the magnetic induction field, and $\overleftrightarrow{\varepsilon}[\omega_{c},\bm{\mathcal{M}}]$ is the magnetization-dependent permittivity tensor. This expression for $u_{\omega_{c}}$ assumes that the frequency of the electromagnetic waves is such that $\omega_{c}\gg\omega_{m}$, where $\omega_{m}$ is the frequency of the magnetization oscillations ($\sim\:{\rm GHz}$), which allows a rotating wave approximation \citep{Bittencourt_2021_ENZ}. At these frequencies we can consider the permeability tensor as dispersionless.

The permittivity tensor has components $\varepsilon_{ij}[\omega,\bm{\mathcal{M}}]=\varepsilon[\omega]\delta_{ij}+i\varepsilon_{ijk}\mathcal{F}[\omega]\mathcal{M}_{k}$ which describes the Faraday effect, a first-order response in the magnetization, in its off-diagonal components. The functions $\varepsilon[\omega]$ and $\mathcal{F}[\omega]$ depend on the specific model for dispersion. The Faraday rotation angle per length $\Phi[\omega]$ quantifies the rotation of the polarization of a linearly polarized electromagnetic wave propagating along $\bm{e}_{z}$ and is given by
\begin{equation}
\Phi[\omega]=-\frac{\omega}{c}\frac{\mathcal{F}[\omega]\mathcal{M}_{S}}{2\bar{n}[\omega]}\label{eq:FradayRotAngle-1}
\end{equation}
where $2\bar{n}[\omega]=\sqrt{\varepsilon[\omega]+\mathcal{F}[\omega]\mathcal{M}_{S}}+\sqrt{\varepsilon[\omega]-\mathcal{F}[\omega] \mathcal{M}_{S}}$ is the average refractive index.

We decompose the electric field as $\bm{E}^{(+)}(\bm{r},t)=\sum_{l}C_{l}\alpha_{l}(0)e^{-i\omega_{c}t}\bm{F}_{l}(\bm{r})$, where $l$ labels the modes and their polarization, $\alpha_l$ the corresponding mode amplitudes, and $C_l$ the field normalization. $\bm{F}_{l}(\bm{r})$ are magnetization-depedent mode functions satisfying $\nabla\times\nabla\times\bm{F}_{l}(\bm{r})-\frac{\omega_{c}^{2}}{c^{2}}\overleftrightarrow{\varepsilon}[\omega_{c},\bm{\mathcal{M}}]\cdot\bm{F}_{l}(\bm{r})=0$,
$\bm{\nabla}\cdot\bm{F}_{l}(\bm{r})=0$, plus appropriate boundary conditions.  For example, in a spherical cavity, those correspond to non-degenerate whispering gallery modes \cite{Ford_1978_scattering_absorption,Almpanis_2020_Spherical_optomagnonic}. After integrating over space, the total energy $U_{\omega_{c}}=\int d^{3}\bm{r}u_{\omega_{c}}$ splits into diagonal (${\rm D}$) and non-diagonal (${\rm ND}$) terms in mode-space,
\begin{equation}
U_{\omega_{c}}=U_{\omega_{c}}^{({\rm D})}+U_{\omega_{c}}^{({\rm ND})},\label{eq:TotEnergy}
\end{equation}
with $U_{\omega_{c}}^{({\rm D})}=\frac{1}{4}\sum_{l}\vert C_{l}\vert^{2}\vert\alpha_{l}\vert^{2}\mathcal{I}_{ll}[\omega_{c},\bm{\mathcal{M}}]$ and $U_{\omega_{c}}^{({\rm ND})}=\frac{1}{4}\sum_{l\neq l^{\prime}}\alpha_{l}^{*}\alpha_{l^{\prime}}C_{l}^{*}C_{l^{\prime}}\mathcal{I}_{ll^{\prime}}[\omega_{c},\bm{\mathcal{M}}]$. While $U_{\omega_{c}}^{({\rm D})}$ yields upon quantization the usual harmonic oscillator Hamiltonian for the photon modes, $U_{\omega_{c}}^{({\rm ND})}$ results in coupling between different modes mediated by the magnetization. The functions $\mathcal{I}_{ll^{\prime}}[\omega_{c},\bm{\mathcal{M}}]$ are given by
\begin{align}
\mathcal{I}_{ll^{\prime}}[\omega_{c},\bm{\mathcal{M}}] & =\int d^{3}\bm{r}\Big\{\varepsilon_{0}\bm{F}_{l}^{*}\cdot\partial_{\omega_{c}}\left(\omega_{c}\overleftrightarrow{\varepsilon}[\omega_{c},\bm{\mathcal{M}}]\right)\cdot\bm{F}_{l^{\prime}} \nonumber \\
 & \,\,\,\,\,\,\,\,\,\,+\frac{1}{\mu_{0}\omega_{c}^{2}}\left(\nabla\times\bm{F}_{l}^{*}\right)\cdot\left(\nabla\times\bm{F}_{l^{\prime}}\right)\Big\}.
\label{eq:OverIntegral}
\end{align}

We work with plane wave-like modes, $\bm{F}_{l}(\bm{r})=e^{i\bm{k}_{l}[\omega_{c}]\cdot\bm{r}}\bm{f}_{l}/\sqrt{V}$, where $V$ corresponds to magnetic volume where the plane waves propagate. The wave equation yields the Fresnel equation $\left[\overleftrightarrow{\mathcal{K}}\cdot\overleftrightarrow{\mathcal{K}}+\frac{\omega^{2}}{c^{2}}\overleftrightarrow{\varepsilon}[\omega_{c},\bm{\mathcal{M}}]\right]\cdot\bm{\bm{f}}_{l}=0$, where $\mathcal{K}_{ij}=-\epsilon_{ijk}k_{l,k}$. Since we are interested in the coupling between photons and magnetic excitations, we consider small fluctuations of the magnetization around its saturation value, assumed to be along the $\bm{e}_{z}$ direction: $\bm{\mathcal{M}}=\mathcal{M}_{z}\bm{e}_{z}+\mathcal{M}_{x}\bm{e}_{x}+\mathcal{M}_{y}\bm{e}_{y}$, where $\mathcal{M}_{x,y}/\mathcal{M}_{z}\ll1$, and we compute all required quantities for quantization, up to linear order in $\mathcal{M}_{x,y}/\mathcal{M}_{z}$. For simplicity we consider an uniformly precessing magnetization (the Kittel mode). In the Voigt configuration, the wave-vector is perpendicular to the saturation magnetization $\bm{k}_{l}\parallel\bm{e}_{x}$, and the wave-vectors obtained from the Fresnel's equation are
\begin{equation}
k_{+}^{2}[\omega_{c}]=\frac{\omega_{c}^{2}}{c^{2}} \varepsilon[\omega_{c}], \quad k_{-}^{2}[\omega_{c}]=\frac{\omega_{c}^{2}}{c^{2}}\left( \varepsilon[\omega_{c}]-\frac{\mathcal{F}^2[\omega_c] \mathcal{M}_S^2}{\varepsilon[\omega_c]}\right),\label{eq:ModeDisp}
\end{equation}
with the magnetization-dependent mode vectors
\begin{align}
\bm{f}_{+} & =\frac{\mathcal{M}_{x}}{\mathcal{M}_{S}}\bm{e}_{x}+\left(\frac{i\varepsilon[\omega_{c}]\mathcal{M}_{x}}{\mathcal{F}[\omega_{c}]\mathcal{M}_{S}^{2}}+\frac{\mathcal{M}_{y}}{\mathcal{M}_{S}}\right)\bm{e}_{y}+\bm{e}_{z}\nonumber \\
\bm{f}_{-} & =-i\frac{\mathcal{F}[\omega_{c}]}{\varepsilon[\omega_{c}]}\mathcal{M}_{S}\bm{e}_{x}+\bm{e}_{y}+\left(\frac{i\varepsilon[\omega_{c}]\mathcal{M}_{x}}{\mathcal{F}[\omega_{c}]\mathcal{M}_{S}^{2}}-\frac{\mathcal{M}_{y}}{\mathcal{M}_{S}}\right)\bm{e}_{z},
\end{align}
where we used $\mathcal{M}_{z}\sim\mathcal{M}_{S}$. These results are valid for homogeneous and unbounded media, or for a Fabry-P\'{e}rot cavity configuration, where the boundary conditions impose discrete wave vectors $k_{\pm}[\omega_{c}]=2\pi n/L$ where $L$ is the cavity length.

In order to quantize Eq.~(\ref{eq:TotEnergy}), we first fix $C_{l}$ such that $U_{\omega_{c}}^{({\rm D})}=\frac{1}{2}\sum_{l}\vert\alpha_{l}\vert^{2}$, and then promote the field amplitudes to bosonic operators via $\alpha_{l}\rightarrow\sqrt{2\hbar\omega_{c}}\hat{a}_{l}$, yielding $U_{\omega_{c}}^{({\rm D})}\rightarrow\hat{H}{}^{({\rm D})}=\sum_{l}\hbar\omega_{c}\hat{a}_{l}^{\dagger}\hat{a}_{l}$ plus a zero-point energy term. The normalization constants are functions of the saturation magnetization $C_{m}=\sqrt{2/ \mathcal{I}_{mm}[\omega_{c},\mathcal{M}_{S}]}$,
given by
\begin{equation}
\begin{aligned}
C_{+}[\omega_{c}]&=\left[\varepsilon_{0}\left(\partial_{\omega_{c}}\left(\omega_{c}\varepsilon[\omega_{c}]\right)+\varepsilon[\omega_{c}]\right)/2\right]^{-1/2} \\
C_{-}[\omega_{c}] & =\Big\{\varepsilon_{0}\Big[\left(1+\frac{\mathcal{F}^{2}[\omega_{c}]\mathcal{M}_{S}^{2}}{\varepsilon^{2}[\omega_{c}]}\right)\partial_{\omega_{c}}(\omega_{c}\varepsilon[\omega_{c}])\\
 & \,\,-2\frac{\mathcal{F}[\omega_{c}]}{\varepsilon[\omega_{c}]}\partial_{\omega_{c}}(\omega_{c}\mathcal{F}[\omega_{c}])\mathcal{M}_{S}^{2}\\
 & +\vert\varepsilon[\omega_{c}]-\frac{\mathcal{F}^{2}[\omega_{c}]\mathcal{M}_{S}^{2}}{\varepsilon[\omega_{c}]}\vert\Big]/2\Big\}^{-1/2}.
 \end{aligned}
\end{equation}
This quantization procedure for the field in a dispersive medium incorporates dispersion in a Hamiltonian formulation \citep{Raymer_2020_Quantum_theory} and reproduces results from Lagrangian-based approaches \citep{Barnett_1996_Field_Commutation,Huttner_1992_Dispersion_and_Loss} for frequencies far away from any absorption resonance. It has been applied to describe emission in negative index media \citep{Milonni_2003_quantized_field} and perturbative and non-perturbative dynamics in the NZI regime \citep{Lobet_2020_Fundamental, Liberal_2021_nonperturbative}.

The optomagnonic Hamiltonian is obtained from $U_{\omega_{c}}^{({\rm ND})}$. For the optical modes, we use the aforementioned quantization, while for the magnetic excitation we consider the Holstein-Primakoff transformation \citep{Holstein_1940_Field}, up to first order in the magnon operators: $\mathcal{M}_{x}\rightarrow\mathcal{M}_{{\rm ZPF}}(\hat{m}^{\dagger}+\hat{m})$, where $\hat{m}^{(\dagger)}$ is the magnon annihilation (creation) bosonic operator and  $\mathcal{M}_{{\rm ZPF}}=\left(\gamma\hbar\mathcal{M}_{S}/2V\right)^{1/2}$ describes the zero-point fluctuations of the magnetization with $\gamma$ the gyromagnetic ratio. The Hamiltonian obtained by this procedure $U_{\omega_{c}}^{({\rm ND})}\rightarrow\hat{H}_{{\rm OM}}$ is given by Eq.~(\ref{eq:OMDispersionHamiltonian}) with the frequency-dependent
optomagnonic coupling constant
\begin{align}
g[\omega_{c}] & =\frac{\omega_{c}}{2}\varepsilon_{0}\mathcal{M}_{{\rm ZPF}}\Xi C_{+}[\omega_{c}]C_{-}[\omega_{c}]\label{eq:CoupVoiu}\\
 & \,\times\left[\frac{\mathcal{F}[\omega_{c}]}{\varepsilon[\omega_{c}]}\partial_{\omega_{c}}\left(\omega_{c}\varepsilon[\omega_{c}]\right)-\partial_{\omega_{c}}\left(\omega_{c}\mathcal{F}[\omega_{c}]\right)\right]\nonumber 
\end{align}
where $\Xi$ is a mode-overlap factor. The quantized Hamiltonian of Eq.~(\ref{eq:OMDispersionHamiltonian}) describes the usual parametric coupling between two photon modes and the magnon quadrature \citep{Kusminskiy_2016_Coupled,Liu_2016_Optomagnonics}, but including dispersion. The coupling Eq.~\eqref{eq:CoupVoiu} is valid for degenerate optical modes at frequencies such that $\varepsilon[\omega]\ge0$. Generalizations are discussed in \citep{Bittencourt_2021_ENZ}. For a frequency $\omega_{{\rm ENZ}}$ at which $\varepsilon\rightarrow0$,
Eq.~(\ref{eq:CoupVoiu}) yields Eq.~(\ref{eq:GEZ}), which is independent from the dispersion's model, as long as $\mathcal{F}^{2}[\omega_{{\rm ENZ}}]$ and $\partial_{\omega}(\mathcal{F}^{2}[\omega])\vert_{\omega_{{\rm ENZ}}}$ do not diverge \citep{Bittencourt_2021_ENZ}. The specific dispersion model defines the value of $\omega_{{\rm ENZ}}$ and for large zero-point fluctuations of the magnetization, $g[\omega_{{\rm ENZ}}]$ can be of the order of the magnon frequency, putting such ENZ optomagnonic systems in the single-magnon
strong coupling regime. Equations (\ref{eq:CoupVoiu}) and (\ref{eq:GEZ}) are general and only assume the plane wave-like modes for the waves in the Voigt configuration. The Cotton-Mouton effect can be included taking into consideration quadratic magnetization terms in the permittivity tensor. Those can generate interference effects that modify the coupling \cite{Haigh_2021_Polarization_dependent}.

In the following we consider a specific model of dispersion. In dielectrics, the Faraday effect is due to electric-dipole transitions between a group of ground states and a group of excited states with broken degeneracy due to perturbations included in the microscopic Hamiltonian, e.g., spin-orbit coupling, crystalline field and Zeeman interaction \citep{Crossley_faraday_1969,Pershan_magnetooptical_1967,Fleury_scatteringoflight_1968}. In this minimal model, the dispersion of the permittivity tensor is given by a Lorentz-like model for which
\begin{equation}
\begin{aligned}
\varepsilon[\omega] & =1+\frac{\omega_{0}^{2}\left(\bar{\varepsilon}-1\right)}{\left(\omega_{0}^{2}-\omega^{2}\right)-i\eta\omega},\label{eq:Permittivity}\\
\mathcal{F}[\omega] & =\frac{A_{3}\omega\omega_{0}}{\left(\omega_{0}^{2}-\omega^{2}-i\eta\omega\right)^{2}},
\end{aligned}
\end{equation}
where $\omega_{0}$ is the resonance frequency for the ionic transitions in absence of perturbations, and $\bar{\varepsilon}$ is the permittivity
for $\omega\ll\omega_{0}$. The factor $\eta$ takes into account absorption and is required by causality. Absorption and losses can
hinder effects typical to the NZI regime, a drawback prominent in metallic NZI systems \citep{khurgin_how_deal_2015,kinsey_near_zero_index_2019} that is mitigated in dielectric platforms \citep{Jahani_2016_alldielectric, Dong_2021_ultra_low_loss, Tang_2021_low_loss_ZI}. The quantity $A_{3}$ depends on products of elements of the electric dipole transition matrix and on the spin-orbit coupling. The model assumes that the ground states have zero orbital angular momentum (such as $S$ states). This corresponds, for example, to ${\rm Fe}{}^{+3}$ ions in Yttrium-Iron garnet (YIG), in which the off diagonal terms of the permittivity tensor have their larger contribution due to spin-orbit coupling \citep{Crossley_faraday_1969}. Otherwise, similar dispersion models for the permittivity tensor describes the behavior of magnetized plasmas \cite{Shen_analogof_2019}, doped dielectrics \cite{Vafafard_2020_tunable_optical_and} and gyrotropic layered strucures \cite{Tsakmakidis_2017_breaking_lorentz}. For a low-loss medium with $\eta\ll\omega_{0}$, the permittivity vanishes at $\omega_{{\rm ENZ}}=\omega_{0}\sqrt{\bar{\varepsilon}}(1-\eta^{2}/2\omega_{0}^{2})$.

We take illustrative values of the relevant parameters corresponding to absorption lines of Yttrium-Iron garnet (YIG). YIG exhibits several absorption lines for wavelengths between $400\,{\rm nm}$ and $900\,{\rm nm}$ \citep{scott_absorptionspectra_1974}. The broad and strong absorption lines around $\sim$ $500\,{\rm nm}$ are related to ionic transitions of the octahedrally oriented ${\rm Fe}^{3+}$ ions, which gives the stronger contribution to the Faraday effect in the medium. Thus, we take as the resonance frequency $\omega_{0}=2\pi c/(500\:{\rm nm})\sim2\pi\times600\,{\rm THz}$. The other parameters for YIG are $\bar{\varepsilon}\sim 4.9$  and $A_{3}^{{\rm YIG}}\sim-2.25\times10^{22}\:{\rm rad}^{2}{\rm Hz}^{2}{\rm m}/{\rm A}$ \citep{Crossley_faraday_1969}. In our single-resonance model, the coefficient $\eta$ in Eq.~(\ref{eq:Permittivity}) corresponding to the absorption coefficient of YIG at $1.2\,\mu{\rm m}$ \cite{scott_absorptionspectra_1974,Stancil_spin_2009} is $\eta/\omega_{0}\sim10^{-6}$. Therefore, for frequencies close to absorption can be safely disregarded and from now on we discard the term $\propto\eta^{2}/\omega_{0}^{2}\sim10^{-12}$ when evaluating $\omega_{\rm{ENZ}}$. In this illustrative framework, the adopted value for the absorption coefficient $\eta$ is very optimistic, in typically in polar dielectrics for ENZ applications $\eta/\omega_{0} \sim 0.03-0.1$ \cite{jahani_all_dielectric_2016}.

\begin{figure}[H]
\begin{centering}
\includegraphics[width=1\columnwidth]{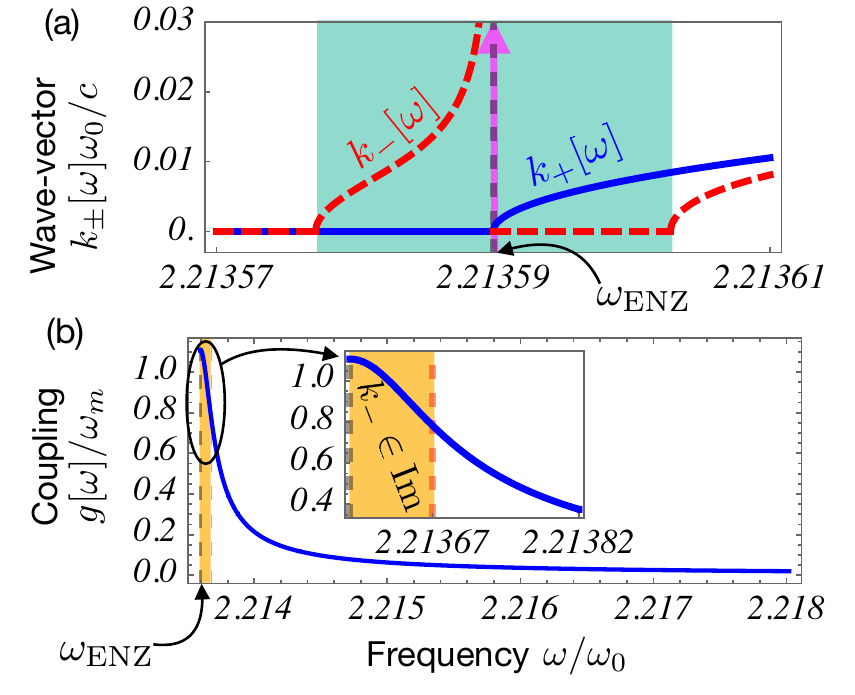}
\par\end{centering}
\caption{(a) Wave vectors of the plane wave modes $k_{\pm}[\omega]$ given by Eq.~(\ref{eq:ModeDisp}) in the Voigt configuration as a function of the frequency close to ENZ point. In a frequency range close to ENZ point, only one of modes propagates; (b) Corresponding optomagnonic coupling given by Eq.~(\ref{eq:CoupVoiu}) as a function of the frequency close to the ENZ point. The inset depicts a zoom-in around $\omega_{{\rm ENZ}}$, showing the coupling enhancement. Illustrative parameters for YIG as discussed in the text and in correspondence with Fig.~\ref{Fig01}. Frequency in units of the ionic transition frequency $\omega_0= 2 \pi \times 600 \, \rm{THz}$, wave vectors in units of $\omega_0/c$, and coupling in units of the magnon frequency $\omega_m=2 \pi \times 10\, \rm{GHz}$.}
\label{Fig02}
\end{figure}

In Fig.~\ref{Fig02} we show the wave vectors $k_{\pm}$ from Eq.~\eqref{eq:ModeDisp} and the optomagnonic coupling $g[\omega_{c}]$ as a function of the photon frequency, for frequencies close to the ENZ frequency. We assume a magnetic volume $\sim(\mu{\rm m})^{3}$ and perfect mode overlap $\Xi=1$. For frequencies smaller than $\omega_{{\rm ENZ}}$, $k_{+}$ is purely imaginary for $\varepsilon<0$ and thus does not propagate, while $k_{-}$ has a real part near $\omega_{{\rm ENZ}}$. For $\varepsilon >0$ and frequencies larger than the ENZ frequency, there is a region were $k_-$ is purely imaginary. This can be exploited for ENZ-based perfect optical isolation \citep{davoyan_optical_isolation_2013}. We notice a divergence of $k_{-}$ at $\omega_{{\rm ENZ}}$ due to the term $\propto\mathcal{F}^{2}/\varepsilon$. For the illustrative values considered in this work, Eq.~\eqref{eq:GEZ} gives $g[\omega_{{\rm ENZ}}]\sim2\pi\times10\,{\rm GHz}$, a value 5 orders of magnitude larger than the optimal theoretical coupling in the near infrarred \citep{Kusminskiy_2016_Coupled} and 9 orders of magnitude larger than the state-of-the-art optomagnonic systems $g\sim50\,{\rm Hz}$ \citep{Zhu_2021_inverse,Haigh_2020_subpicoliter}. 

The magnon-photon coupling close to $\omega_{{\rm ENZ}}$ is comparable to the magnon frequency $g[\omega_{{\rm ENZ}}]\sim\omega_{m}$ for the dispersion model we consider, putting the system in the single-magnon strong coupling regime. In general, for a given medium, $\frac{\mathcal{M}_{{\rm ZPF}}}{\mathcal{M}_{S}}\omega_{{\rm ENZ}}>\omega_{m}$ imposes a constraint on the volume. Using a cavity to confine the modes to small volumes therefore serves to boost the value of the coupling. Such effects can be probed in a driven cavity system via the power spectrum of the light, which exhibit resonance peaks and a frequency shift not present for weak coupling \cite{Rabl_2011_Photon_Blockade, Nunnenkamp_2011_Single_Photon}. To analyze such fingerprints, we consider the Hamiltonian
\begin{align}
\frac{\hat{H}}{\hbar} & =\omega_{c}\left(\hat{a}_{+}^{\dagger}\hat{a}_{+}+\hat{a}_{-}^{\dagger}\hat{a}_{-}\right)+\omega_{m}\hat{m}^{\dagger}\hat{m}\nonumber \\
 & +i g[\omega_{c}]\left(\hat{a}_{+}^{\dagger}\hat{a}_{-}-\hat{a}_{-}^{\dagger}\hat{a}_{+}\right)\left(\hat{m}^{\dagger}+\hat{m}\right).\label{eq:FullHamiltonian}
\end{align}
The eigenenergies of Eq.~(\ref{eq:FullHamiltonian}) are obtained by a change of basis for the photon modes, defining the degenerate modes $\hat{a}_{1,2}$, and a two-mode polaron transformation yields the non-linear energy spectrum $E_{n_{1},n_{2},n_{m}}/\hbar=\omega(n_{1}+n_{2})-\frac{g^{2}}{\omega_{m}}(n_{1}-n_{2})^{2}+\omega_{m}n_{m}$, where $n_{1,2,m}$ are positive integers \cite{Bittencourt_2021_ENZ}.


We assume that mode $\hat{a}_{1}$ is coherently driven, and consider its spectrum $S_{1}[\omega]=\int dt\langle\hat{a}_{1}(t)\hat{a}_{1}^{\dagger}(0)\rangle e^{-i\omega t}$. In the single-magnon strong coupling regime, linearization of the Hamiltonian Eq.~\eqref{eq:FullHamiltonian} is not adequate to completely describe the system, thus to obtain $S_{1}[\omega]$ we resort to numerical simulations of the Lindblad master equation for the density matrix $\rho$ of the system 
\begin{equation}
\dot{\rho} =\frac{i}{\hbar}[\rho,\hat{H}+\hat{H}_{{\rm Dr}}]+\sum_{i=\pm}\kappa_{i}\mathcal{D}[\hat{a}_{i}]\rho+\gamma\mathcal{D}[\hat{m}]\rho,
\end{equation}
where the dissipation superoperator is $\mathcal{D}[\hat{O}]\rho=\hat{O}\rho\hat{O}^{\dagger}-\left\{ \hat{O}^{\dagger}\hat{O},\rho\right\} /2$. We have assumed that the photons and the magnon baths are at zero temperature, and added the coherent driving of mode $1$: $\hat{H}_{{\rm Dr}}=i\hbar\Omega(\hat{a}_{1}e^{i\omega_{{\rm D}}t}-\hat{a}_{1}^{\dagger}e^{-i\omega_{{\rm D}}t})$, where $\Omega$ is the driving amplitude and $\omega_{{\rm D}}$ is the driving frequency, which we set to $\omega_{c}$ (mode 1 is driven at resonance). We focus our attention on a weakly driven system, $\Omega\ll\omega_{m}$, to keep the size of the Hilbert space used in the simulations tractable
\citep{Johansson_2013_qutip}.

\begin{figure}
\centering{}\includegraphics[width=1\columnwidth]{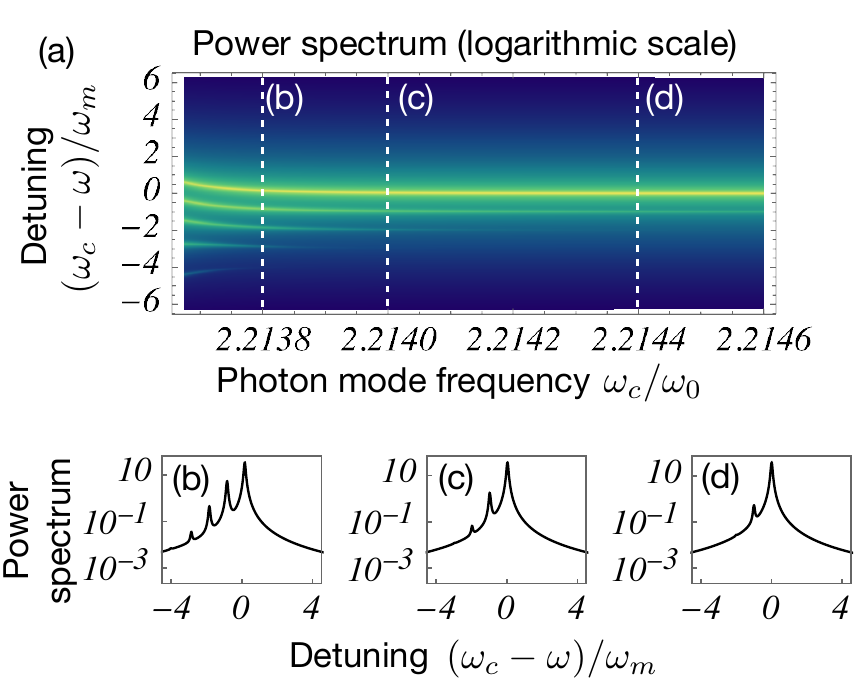}\caption{Power spectrum  as a function of the mode frequency $\omega_{c}$ and as a function of the frequency $\omega$. We work in a frame rotating at the driving frequency. Several peaks equally spaced by $\sim\omega_{m}$ are present when the coupling is comparable to the optical decay rate and to the magnon's frequency, and the resonance lines exhibit a characteristic shift close to the ENZ frequency due to the non-linear energy spectrum. Parameters: $\kappa_{1}/\omega_{m}=10^{-1}$, $\Omega/\omega_{m}=10^{-4}$, $\gamma/\omega_{m}=10^{-3}$. Detuning in units of the magnon mode frequency $\omega_m = 2 \pi \times 10 \, \rm{GHz}$ and photon mode frequency in units of the ionic transition frequency $\omega_0= 2 \pi \times 600 \, \rm{THz}$.}
\label{Fig03}
\end{figure}

The power spectrum $S_{1}[\omega]$ is depicted in Fig.~\ref{Fig03} as a function of the frequency $\omega_{c}-\omega$, in the frequency
range in which both optical modes have real wave-vectors, and for an optical decay rate $\kappa_{1}=2\pi\times1\:{\rm GHz}$. The several peaks observed for $\omega_{c}\sim\omega_{{\rm ENZ}}$, a regime in which both $g\sim\omega_{m}$ and $g>\kappa$, are fingerprints of the strong magnon--photon coupling not captured in a linearzed model. The first peak occurs at $\omega=\omega_{c}-g^{2}/\omega_{m}$ and corresponds to a transition from 0 to 1 photon state with frequency shifted by the strong optomagnonic coupling. The other peaks are equally spaced by about $\sim\omega_{m}$ and correspond to processes creating magnons. Such behavior for $S_{1}[\omega]$ at zero temperature is exclusively due to quantum effects of the single magnon strong coupling regime. Since we are assuming zero temperature baths, a photon can only create a magnon and thus $S_1$ only display peaks at frequencies $>\omega_{c}-g^{2}/\omega_{m}$. At finite temperature, peaks at frequencies $<\omega_{c}-g^{2}/\omega_{m}$ would appear, indicating a finite probability for the absorption of a magnon by a photon.

The features displayed by the power spectrum depend on the cavity quality factor. If the optical mode decay rate is $\kappa_{1}>g[\omega_{c}]$, the peaks displayed by $S_1$ would be suppressed by the large cavity linewidth. Such decay rate includes both radiative decay and intrinsic decay, related to absorption. As discussed above, in our model and for our illustrative parameters, absorption is small close to the ENZ frequency and was disregarded in the quantization procedure. For bulk media, the additional resonances related to other ionic transitions introduce more losses, which can hinder the ENZ behavior \citep{khurgin_how_deal_2015}. Therefore, such systems require design and optimization, for example in combination with plasmon-polariton systems and with structured media to minimize absorption. For our parameters, the condition $\kappa_{1}<g[\omega_{{\rm ENZ}}]$ requires quality factors $>10^{6}$ for an optomagnonic device operating at the ENZ frequency with the parameters used in this text. The cavity power spectrum can be measured in a homodyne experiment with the output light field from the cavity. Such measurement would also require the magnon mode to be close to its ground state to minimize thermal noise, requiring temperatures of $\sim 70\,$ mK for magnons with frequency of $10$ GHz, which can be achieved with a dilution fridge. 

To summarize, we have obtained the frequency-dependent optomagnonic coupling, which includes dispersion of the dielectric's permittivity, and we have shown that the optomagnonic coupling close to the ENZ frequency is greatly enhanced, becoming comparable to the magnon's frequency. Optomagnonic strong coupling effects are prominent in systems with a high quality factor, and can be measured via the light power spectrum. Alternative measurements of the strong coupling includes modifications of the magnon induced transparency \citep{Liu_2016_Optomagnonics, Kronwald_2013_Optomechanically_Induced}, and the counting statistics of the optical field \citep{Rabl_2011_Photon_Blockade}. Strong magnon-photon coupling can allow applications such as efficient optical cooling of magnons \citep{Sharma_2018_Optical_Cooling} and  all-optical generation of quantum states of the magnetization \citep{Bittencourt_2019_Magnon_Heralding}. Furthermore, the bulk medium dispersion considered in this work can be combined with geometrical dispersion or with structured media to tailor the ENZ response \citep{liberal_near_zero_2017,kinsey_near_zero_index_2019}. 

\begin{acknowledgments}
The authors thank R. Boyd and O. Reshef for useful discussions and comments on the manuscript. V.A.S.V. Bittencourt and S. Viola Kusminskiy acknowledge financial support from the Max Planck Society. I.L. acknowledges support from ERC Starting Grant 948504, Ram\'{o}n y Cajal fellowship RYC2018-024123-I and project RTI2018-093714-301J-I00 sponsored by MCIU/AEI/FEDER/UE.
\end{acknowledgments}

\bibliographystyle{apsrev4-1} 
\bibliography{Bibliography_Microscopic}

\end{document}